\documentstyle[12pt,psfig]{article}
\begin{document}
\newcommand{\nwc}{\newcommand}
\def\gtrsim{\mathrel{\relax{\raisebox{3pt}{$\mathord{>}$} \kern-0.75em
\raisebox{-2pt}{$\sim$}}}}
\nwc{\lng} {\langle}
\nwc{\rng} {\rangle}
\nwc{\lf} {\left}
\nwc{\ri} {\right}
\nwc{\pr} {^{\prime}}
\nwc{\pa}{\partial}
\nwc{\be}  {\begin{equation}}
\nwc{\ee}  {\end{equation}}
\nwc{\D}  {\displaystyle}
\nwc{\nnn} {\nonumber \vspace{.2cm} \\ }
\nwc{\de} {\delta}
\nwc{\La}  {\Lambda}
\nwc{\Rc}  {{\cal R}}
\nwc{\pro}{\propto}
\nwc{\la} {\lambda}
\nwc{\Oc}  {{\cal O}}

\begin{titlepage}
\setlength{\topmargin}{1cm}
\title{Quantum Inflation? \thanks{Partially supported by
Deutsche Forschungsgemeinschaft and EC grant SC1--CT91--0729.}\vspace{1 cm}}

\author{{{\sc Zygmunt Lalak}
  \thanks{On leave from the Institute of Theoretical Physics,
   University of Warsaw.} \ \ \ and {\sc \ \ Rudolf Poppe}}   \\ \\
       {\em Physik Department} \\
       {\em Technische Universit\"at M\"unchen} \\
       {\em D--85748 Garching, Germany}}

\date{ }
\maketitle

\begin{picture}(0,0)(-350,-300)
\put(12,0){TUM--HEP--207/94}
\put(12,-20){August 1994}
\end{picture}

\thispagestyle{empty}
\vspace{3cm}

\begin{abstract}
We consider curved space quantum corrections to the equations of motion
of the inflaton field in the early Universe. Using the stochastic formalism
in phase space, we demonstrate that the quantum corrected evolution of the
inflaton can differ dramatically from its classical evolution when
the mass scales in the potential become large, which is naturally the case
in fundamental theories describing Planck scale physics. Using the example
of the cosine potential, we show that the prolonged, perhaps even eternal,
quantum--inflationary period is expected with a significant probability. This
feature of Planck scale potentials can be dangerous, but it offers also a
possibility of creating the inflationary phase in string or
supergravity models, where no natural realization of inflation has been
found so far.
\end{abstract}

\end{titlepage}

The cosmological standard model demands an unacceptable amount of
fine tuning concerning the initial conditions of the Universe.
A superluminal expansion of the scale factor, appearing during
an inflationary epoch at a very early stage of the evolution of the
Universe, was thought to circumvent these unnatural features
\cite{guth}. However, the classical
analysis of most of the proposed inflationary models shows that also
inflation demands a fine tuning of either the initial state
(new inflation) or the parameters of the potential (nearly every model).
One of the reasons is the need for flat potentials
coming from the fact that during inflation at least 60 e--folds must be
attained to solve the ``standard'' horizon, flatness and monopole puzzles
\cite{guth}.

In the class of fundamental theories describing Planck scale physics
(superstrings, quantum gravity) the natural scale,
which sets the (nonvanishing)
expectation values and masses of various fields,
is the Planck scale itself, $M_p \approx 10^{19} \, $ GeV.
Hence, one has to face the problem of Planck scale potentials
possibly
playing, as we demonstrate below, a prominent role in the physics of the
very early Universe.
As important examples let us mention the so--called moduli fields arising
in stringy models (they are allowed to have a potential when supersymmetry is
broken), for instance the stringy dilaton which is believed to play an
important role both in cosmology and in low--energy phenomenology and whose
potential,
perturbative or nonperturbative, remains undetermined.

Classical analysis shows that sufficient inflation
is difficult to achieve in the classical potentials
with the mass scales lying at $M_p $.
Since one expects that the influence of quantum effects
on the dynamics of the inflaton
(by inflaton we understand any Planck scale field dominating
the energy density of the Universe) and the
metric is especially significant as the energy density approaches $M_p^4$,
it is important to include quantum--mechanical effects into the analysis.
Attempts have been made to describe the
fluctuating inflaton field during the slow roll phase
by  solving explicitly the Fokker--Planck equation \cite{rey},\cite{salopek2}.
In the stochastic approach, which we will adopt here,
quantum fluctuations of the
inflaton are investigated in phase space.
This method has been  applied to the usual chaotic potentials in \cite{yi}.

In this note we follow the stochastic method and reexamine the evolution
of the inflaton in phase space for the
cosine potential, which has been analysed classically under the name
"natural inflation" \cite{freese},\cite{adams}. We present the results of the
numerical analysis based on stochastic dynamics \cite{yi} with the mass
parameters
of the ``natural'' potential both lying at the Planck scale.
We choose homogeneous universes with varying cosmic kinetic
energies. A minimal coupling between gravity and the inflaton field is assumed.

By means of stochastic dynamics,
quantum fluctuations are given a statistical interpretation
(quantum noise). The method takes into account the fluctuations in the local
expansion rate due to the coupled evolution of the inflaton and the metric.
The inflaton and its first time derivative (the inflaton ``velocity'') are
decomposed into two parts with respect to the physical horizon $H^{-1}$
 ($H$ is the
Hubble parameter):
a long--wavelength (coarse--grained) part and a short--wavelength one.
The quasi--classical field, whose evolution one follows, is formed by the
coarse--grained component,
and the fluctuations arising in all Fourier modes are represented by the
random force (the noise) in its equations of motion.

The coupled evolution equations for the coarse--grained components of
the inflaton $\phi $ and the velocity $v=d\phi /dt$ are given by
\be \D \begin{array}{lcl}
\dot{\phi} (t) & = & v (t)+\D{\frac{H^{3/2}}{\sqrt{8\pi ^2}}}\,\eta (t)
\vspace{4mm} \\
\dot{v} (t) & = & -3Hv (t) - V\pr (\phi ) \vspace{4mm} \\
H^2 & = & \D{\frac{8\pi }{3M_p^2 }}\lf[\D{\frac{1}{2}}\, v^2 (t)
+V(\phi )\ri] \ ,
\label{stoevoleq}
\end{array} \ee
where the prime denotes $d/d\phi \, $.
The first equation has the form of a classical stochastic equation
of motion
(Langevin--equation) \cite{risken}, where the random noise $ \eta (t)$
includes the effects of the rapidly fluctuating Fourier modes.
This random noise term is missing in the classical analysis.
$\eta (t) $ is Gaussian random with zero mean and a correlation function
$\lng\eta (t)\eta (t\pr )\rng =2\,\de (t-t\pr )$,
which shows the Markov property of the process.
In the approximation we work all spatial correlations in the
random force are neglected to make the problem tractable
(also allowing us to interpret ensemble averages as space averages).
The cosine potential is given in terms of the parameters $\La $ and $f$
      \be
      V(\phi)=\La ^4\,\lf[ 1+\cos\lf(\frac{\phi}{f}\ri)\ri] \ .
      \label{cos}
      \ee
For simplicity we  use $\La =f=M_p $, although the analysis stays valid for any
set of $M_p$--magnitude scales.
The coarse--grained parts of the fields, $\phi $ and $v $, are interpreted
as classical random variables, whose evolution is modified by the influence
of the stochastic noise. The average over a large number of independent
realizations  of the trajectories
of the coarse--grained component is considered to be the
classical homogeneous background.
Treating the problem in phase space with the two independent phase space
variables $\phi$ and $v$ has the advantage that initial
conditions with a dominating kinetic energy contribution, $\frac{1}{2}\,
\dot{\phi}^2 \gtrsim V(\phi )$ are allowed. The method makes it also possible
to
control explicitly the fluctuations at the initial and final stages of
inflation.
There exists  a well known procedure of solving a set of Langevin--equations by
computer simulations, the molecular dynamics method \cite{risken}. Although
the accuracy of the method is not very good (it is of the order $\tau$ after
N steps of integration; $\tau$ is the step length), it provides a good
estimate for the behaviour of the mean trajectory in phase space which is
finally compared with the classical evolution.
Usually, It\^o's definition of the stochastic noise is used \cite{risken}.
The discrete evolution equations then take the form
\be \D \begin{array}{lcl}
\phi_{n+1} & = & \phi_n + v_n \,\tau +\D{\frac{H_n^{3/2}}{\sqrt{8\pi^2}}}
                 \, w_n\,\sqrt{2\tau} \nnn
v_{n+1} & = & v_n -3H_n v_n \,\tau -V\pr _n \,\tau \nnn
H^{\, 2}_n & = & \D{\frac{8\pi}{3M_p^2}}\:\Big[\:\D{\frac{1}{2}}\,
\,v^2_n + V_n \:\Big] \ ,
\end{array}
\label{numevoleq}
\ee
where the index $n$ characterizes the $n$th step of integration.
The stochastic noise is simulated by a random number generator producing
uniform deviates. They are turned into Gaussian deviates $w_n $ with
zero mean $\lng w_n \rng =0$ and unit variance $\lng w_n \, w_{n\pr}\rng
=\de_{nn\pr}$ by the Box--Muller method \cite{recipes}. Simultaneously,
the iterations are performed using a different random number at each step.
Finally, the average for a large number of realizations is taken.

\begin{figure}[phtb]
\renewcommand{\baselinestretch}{1.1} \small \normalsize
%    \par\vspace{3mm}
     \centerline{\psfig{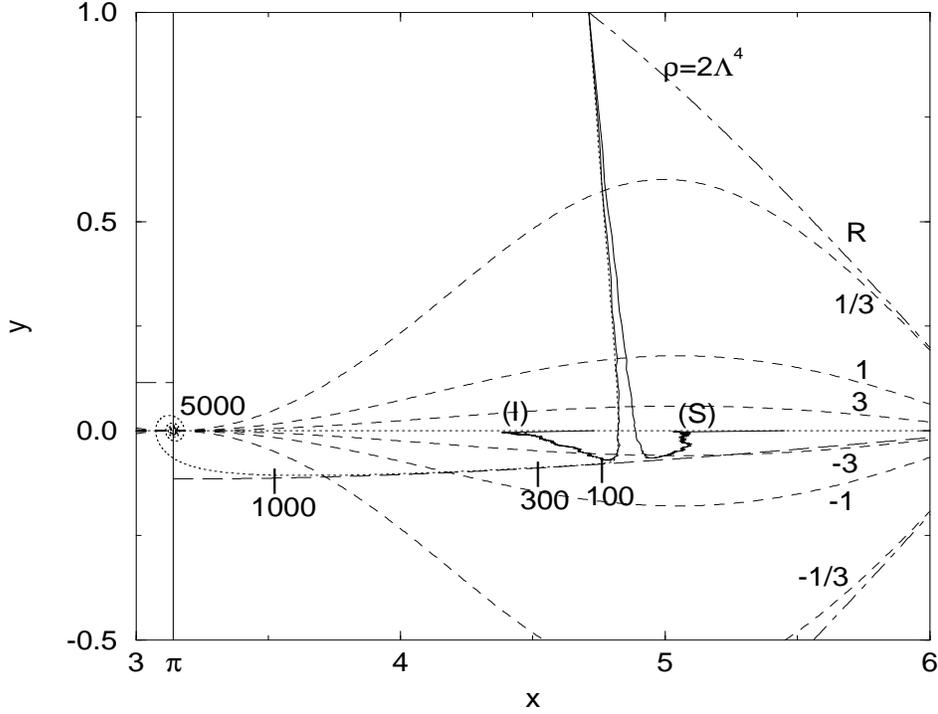}}
%    \par \vspace{2mm}
     \caption{Stochastic mean trajectories (averaged over $3\times 10^4 $
              realizations) in the potential
              $ V(\phi)=\La ^4\,\lf[ 1+\cos\lf(\phi /f \ri)\ri] $
              according to It\^o's (I) and Stratonovich's (S) interpretation
              of the stochastic noise. The parameters $\La $ and $f$
              are set to the Planck scale.
              The classical trajectory is shown by the dotted curve.
              It meets the slow roll curve (long dashed)
              after about 30 steps of integration.
              The initial condition is chosen to lie on the
              curve of constant energy density $\rho =2\,\La^4 $ (dot--dashed
              lines) with a field value of $\phi /f =3\pi /2 $ and positive
              velocity. $3\times 10^4 $ integration steps were performed with
              a step size of $\tau =10^{-2} $. Several lines of
              constant ratio $\Rc $ (dashed lines) are shown.
              The absolute value of $\Rc $ increases,
              if one moves towards $y=0$. The values of $\Rc $ and the number
              of integration steps along the classical trajectory are
              indicated. }
     \label{fig1}
     \par \renewcommand{\baselinestretch}{1.5}
\end{figure}

The choice of a different interpretation of the stochastic noise is usually
expected to have only a small influence on the final result \cite{yi}.
Nevertheless, we perform the integration according to Stratonovich's rule to
show that deviations are possible. In the Stratonovich
case the deterministic drift term $v$ in the first equation of the system
(\ref{stoevoleq}) acquires an additional contribution, the noise--induced
drift:
\be
D^{(1)}_{noise} = \D{ \frac{1}{2}\,\frac{\pa}{\pa\phi}\,
D^{(2)}_{\phi\phi} } = \D{\frac{1}{2}\,\frac{\pa}{\pa\phi}\,
      \frac{H^{3}}{8\pi^2}}
= \D{\frac{3}{16\pi^2}\,H^2 \,\frac{\pa H}{\pa\phi}} \ .
\label{stratdiff}\ee
$D^{(2)}_{\phi\phi} $ is the only nonvanishing entry of the
$2\times 2 $--diffusion matrix. Using the Stratonovich rule,
the first equation of (\ref{numevoleq}) has to be replaced by
\be
\phi_{n+1} =\phi_n + v_n \, \tau + D^{(1)}_{noise,\, n}\,\tau
           +\frac{H_n^{3/2}}{\sqrt{8\pi^2}}\, w_n \sqrt{2\tau} \ .
\label{stratnum}\ee
We present the results of the numerical integration of the system
(\ref{numevoleq}) in the dimensionless variables
\be
x\equiv\frac{\phi}{\La} \ ,\hspace{5mm}
y\equiv\frac{v}{\sqrt{2}\,\La^2} \ ,\hspace{5mm}
\tau\equiv t \La
\label{dimless3}
\ee
for the cosine potential given by (\ref{cos}).
In Fig. \ref{fig1} the evolution of the stochastic mean
trajectories (according to
It\^o's (I) and Stratonovich's (S) interpretation of the stochastic noise)
are compared with the classical evolution of the inflaton in phase space.
In the example we present here,
the initial conditions are  chosen in such a way
that the total energy density
of the inflaton is $2\, \La^{4} $ and the field is positioned at
$\phi /f =3\pi /2$.
This initial value of the field is far away from the region around
the maximum of the potential, where, by some fine tuning, one can
enhance the number of e--folds in the classical case. No fine
tuning of the initial conditions is required in our case.
The mean trajectories shown are averages over $3\times 10^4 $
independent stochastic trajectories. They share  three
characteristic features: (i) significant deviations from the classical path
occur close to the point where, they approach  the classical slow roll
trajectory, (ii) the mean velocity tends toward zero before the oscillating
regime is reached,
(iii) the averaged phase space evolution clearly depends upon the
interpretation of the stochastic noise. Stratonovich's definition (S)
prefers higher energies (this feature agrees with some earlier observations
 \cite{salopek2}). The
general  behaviour of the two mean trajectories is however similar.

The ratio of the noise--induced drift to the classical drift,
\be
\Rc\equiv\frac{D^{(1)}_{noise} }{D^{(1)}_{class.} }=
\D{\frac{3}{16\pi^2 \, v}\,H^2 \,\frac{\pa H}{\pa\phi}}=
\frac{1}{\sqrt{6\pi} \, M_p^3 \, v}\,\sqrt{\frac{1}{2}\, v^2
+V(\phi ) }\; \frac{dV(\phi )}{d\phi }
\label{ratio}
\ee
is also shown in Fig. \ref{fig1}. Here the equation (\ref{stratdiff}) has been
used together
with the equality $D^{(1)}_{class.} =v$.
The noise--induced drift acts always opposite
to the classical one during the stage of the classical slow roll, so that
the {\em averaged} evolution is being
decelerated until the mean velocity vanishes.

\begin{figure}[phtb]
\renewcommand{\baselinestretch}{1.1} \small \normalsize
%    \par\vspace{3mm}
     \centerline{\psfig{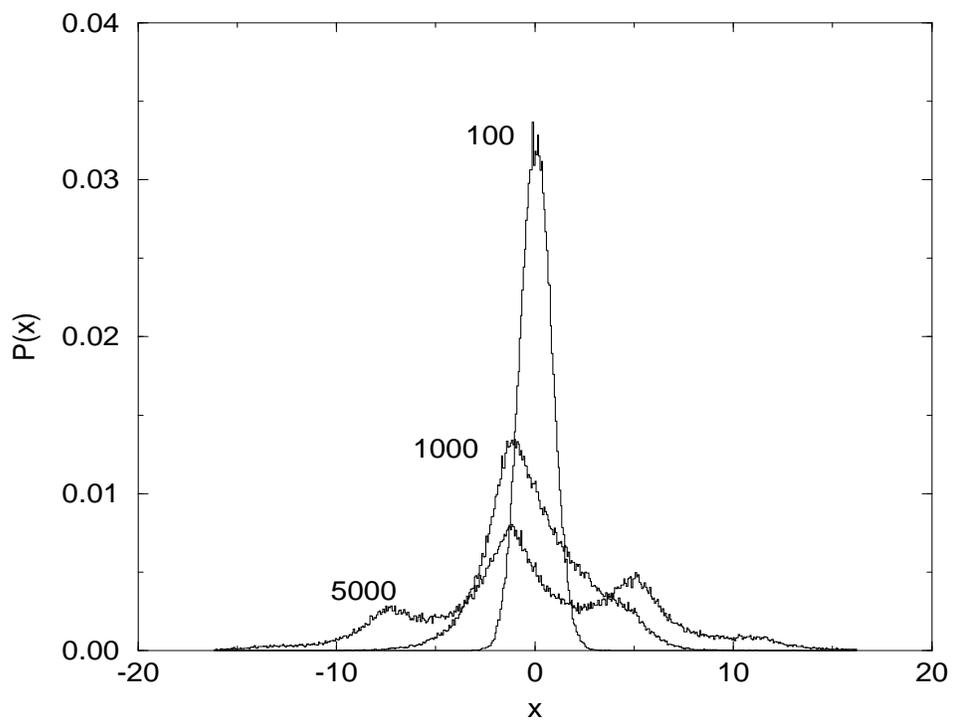}}
%    \par \vspace{2mm}
     \caption{Distribution of the $\phi$--values of $10^{5} $
       realizations (normalized) around the mean (x=0) after 100, 1000
       and 5000 steps of integration. The influence of the potential is
       visible. }
     \label{fig2}
     \par \renewcommand{\baselinestretch}{1.5}
\end{figure}

\begin{figure}[phtb]
\renewcommand{\baselinestretch}{1.1} \small \normalsize
%    \par\vspace{3mm}
     \centerline{\psfig{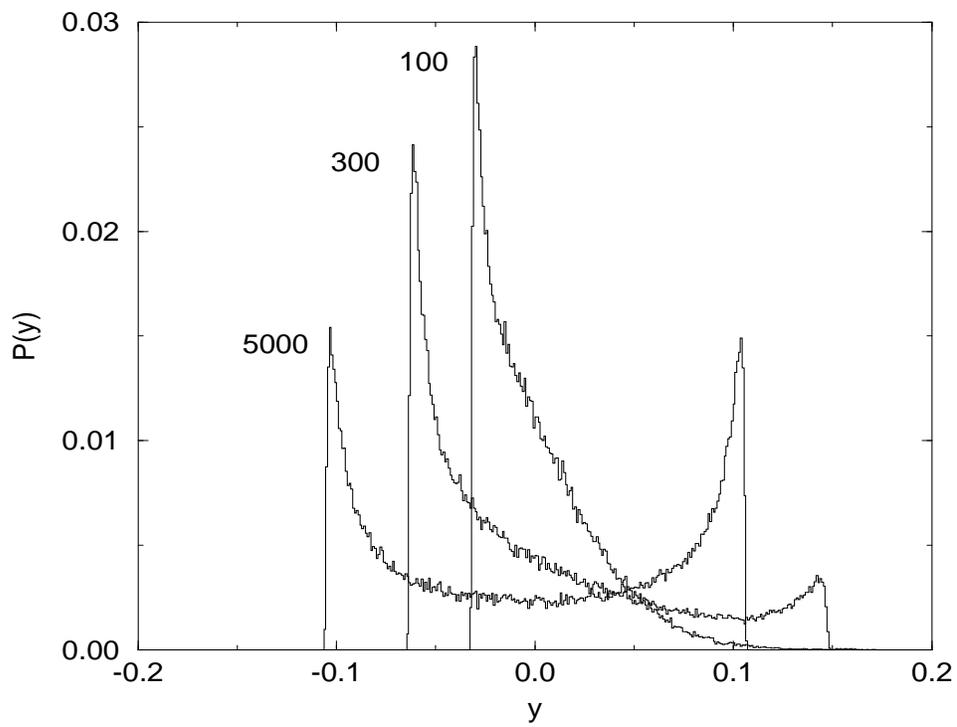}}
%    \par \vspace{2mm}
     \caption{Distribution of the $v$--values of $10^{5} $
       realizations (normalized) around the mean (x=0) after 100, 300
       and 5000 steps of integration. }
     \label{fig3}
     \par \renewcommand{\baselinestretch}{1.5}
\end{figure}

The averaged quantities are not the whole story. One has to
consider the distribution of the individual realizations around the mean.
This is shown in Fig. \ref{fig2} and Fig. \ref{fig3} for the
field and the corresponding
velocity according to the It\^o's rule. The numbers of integration
steps are indicated. The step length was chosen to be $\tau =10^{-2} $, so that
100 steps are equivalent to one Planck time $M_{p}^{-1} $.
The steps are also given in
Fig. \ref{fig1} for the classical trajectory.

At the beginning, $t\simeq\Oc (M_{p}^{-1} )$, when
diffusion dominates over other dynamical effects,
the distribution of the $\phi $--values around the mean remains Gaussian
in its shape.
Gradually, the $\phi$--dependent (nonlinear) drift coefficients
cause deviations from the Gaussian distribution, which get significant
at the time when the classical evolution would have already entered the
epoch of coherent oscillations. It is clearly visible that at this stage the
field has acquired values ranging over several minima of the potential.
The maxima of the distribution are located near the
minima of the potential. Diffusive processes are able to make the field
drop into several different vacua.

The $v$--distribution is also purely Gaussian at the beginning although this
is not shown in Fig. \ref{fig3}. The nonlinearity of the potential
gives rise to significant deviations from the Gaussian shape already long
before the initial velocity has decayed.
It is remarkable that the distribution of this variable
differs from Gaussian statistics considerably earlier than that of the
$\phi $--variable.
Gradually, it develops a tail toward positive velocities.
At this stage the mean velocity constantly decreases,
whereas the tail forms a second maximum, so that the distribution becomes
symmetric with respect to the mean.
This is not unexpected since the quantum "kicks"
appear with similar probability in  opposite directions
(in the $\phi $--space), so that
upward and  downward motions are generated in equal abundances.
One should note that the
total width of the $v$--distribution is much smaller than that of the
$\phi $--distribution.

The final stage of the stochastic evolution
is characterized by a zero mean velocity. This is due to the symmetry of
the $v$--distribution.
The averaged
drift is zero, so that the averaged field seems to stick to one value.
However, the most probable values of the velocity are by no means zero (thus
indicating the dominance of the fluctuations in the inflaton dynamics);
positive velocities are then equally probable as negative velocities, so
that the noise--induced movement uphill has nearly the same probability as
the downward motion.
We have found differences between the time evolution of the inflaton
for It\^o's and Stratonovich's interpretations of the stochastic noise.
But both cases share the same basic feature: since the
noise--induced drift always acts against the classical drift,
the time evolution
of the mean quasi--classical field (compared with the classical case)
can be delayed by quantum fluctuations.
The stochastic analysis shows
that the duration of the quasi--de Sitter phase can be prolonged to reach
a large number of e--folds, much larger than 60.
The
{\em a priori} probability
for this to happen is estimated to be of the order of one
(see discussion below).

The behaviour described so far is generic for any Planck scale field
dominating the energy density of the Universe. The discussion demonstrates
clearly that the classical description of the cosmological evolution driven by
such a field is highly inadequate and that a prolonged quasi--de Sitter
phase is naturally generated in the quantum description.

Let us try to understand whether this quantum inflation can provide us
with a model for successful inflation, solving cosmological puzzles.
The basic requirement for the scenario under consideration to
 serve as a successful inflationary model is that the number of e--folds
produced during the superluminal stages of the evolution of the Universe
 must be large enough,
larger than 60 at least. Of course, there are model universes in the ensemble
where the inflaton rather rapidly approaches values near the minimum of the
potential and then stays there. In this case the generated number of
e--folds is insufficient.

\begin{figure}[phtb]
\renewcommand{\baselinestretch}{1.1} \small \normalsize
%    \par\vspace{3mm}
     \centerline{\psfig{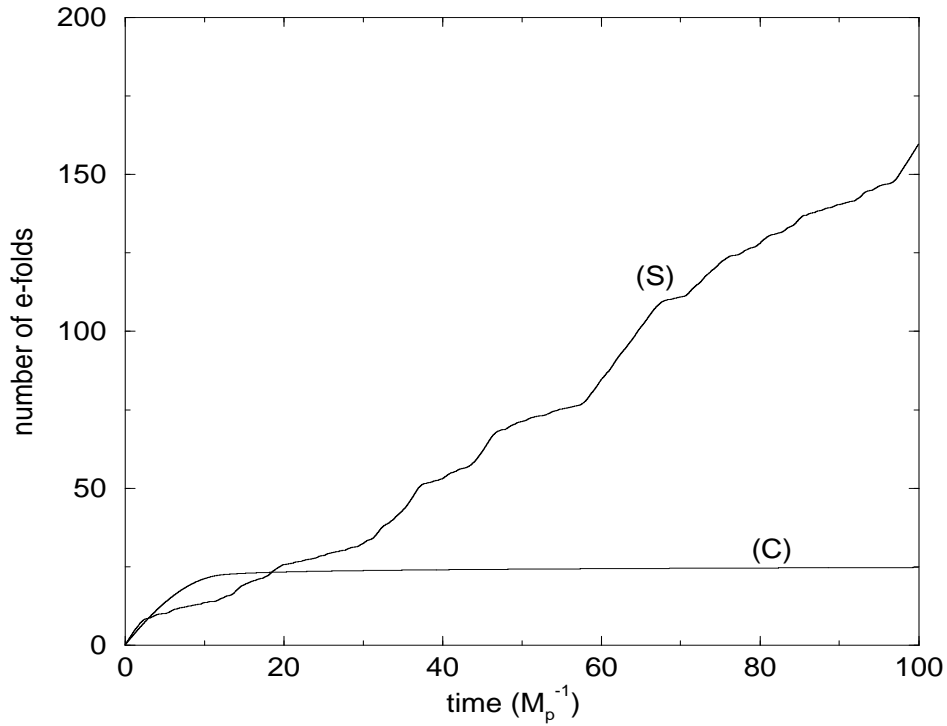}}
%    \par \vspace{2mm}
     \caption{Total number of e--folds produced
              during the evolution of the inflaton. The initial condition
              and the parameters of the potential are chosen as in Fig. 1.
              The classical
              analysis (C) clearly shows the end of inflation after the field
              has reached the minimum of the potential. Only 25 e--folds
              can be produced before the entry into
              the hot Friedmann universe.
              The time behaviour of a
              typical stochastic solution (S) reveals a continious
              superluminal expansion of the universe; the demanded
              number of 60 e--folds is easily exceeded. }
     \label{fig4}
     \par \renewcommand{\baselinestretch}{1.5}
\end{figure}

However, there is a large enough fraction (of order unity)
of all observed realizations of the scenario
where the fluctuations act in such a way that the field varies (over
macroscopic periods of time) much more slowly
than required by the classical evolution. In this case,
a sufficiently large number of e--folds
could be generated to solve the cosmological puzzles.
Fig. \ref{fig4} shows the development of the number of e--folds
produced during the evolution of the inflaton for one typical
stochastic path compared with the classical result. Apparently,
the quantum dominated motion easily provides more than 60 e--folds.
This means that an
inflationary epoch is generated by quantum fluctuations for a generic
inflaton potential with parameters fixed at the
Planck scale. Our results  indicate that
even the {\em a priori} probability of quantum inflation is large,
not to mention the {\em a posteriori} probability \cite{adams},
which is exponentially close to one.
We stress again, that the prolonged
quasi--de Sitter stage is a generic feature of our scenario and
does not require a fine tuning of the initial conditions.

The end of the inflationary period and the reheating of the Universe in this
scenario have
to be examined more carefully. Generally, the strength of the fluctuations
increases
with increasing  values of the potential (as the Hubble parameter increases),
and the field can easily  be driven uphill
thus leading to a prolongation of the quasi--de Sitter phase. Hence, the
possibility of
"eternal" inflation \cite{linde} is incorporated in this model. A natural exit
is always possible with the assumption of the instability of the inflaton.
By decaying into light particles a sufficient
reheating temperature necessary for
baryogenesis could be provided and inflation be stopped.
A specific feature of this violently fluctuating system is
the non--zero
probability of large fluctuations throwing the field out of the
inflationary regime thus ending abruptly the quasi--de Sitter phase.

One possible problem of quantum Planck scale inflation is the
 excessive production of energy density inhomogeneities.
Since the
relative magnitude of a physical length scale to the horizon,
$\la /H^{-1} \pro \dot{R} $, is a fluctuating quantity, any scale $\la $
crosses the horizon not only once, but frequently within the inflationary
period.
For this reason, the resulting amplitude of generated
density perturbations need
to be calculated carefully, as it would serve as a very sensitive probe of the
present scenario.
One obvious way out of the density fluctuations problem is to assume that the
final stage of inflation, when density perturbations at observable scales
are produced, is controlled by some other, lower--scale, field (like in
multiple--inflaton models \cite{linde}),
which can easily be the case in models, like string
inspired ones, where the number of relevant degrees of freedom is
rather large. In such a case, quantum inflation is the natural trigger for
the subsequent inflationary epoch and can easily provide a quasi--de Sitter
stage of realistically long duration. \vspace{5mm}

On the basis of our experience with a wide class of potentials
(we have found similar effects for the quadratic and
quartic potentials of chaotic inflation) we want to conclude that the
presence of quantum inflation seems to be a generic feature of field
theoretical models addressing Planck scale physics and it may be interesting
and fruitful to explore this phenomenon in more detail.

\vspace{15mm}
\centerline{\bf Acknowledgements}
\vskip 0.5cm
We thank Andr\'e Lukas for stimulating discussions,
Stephan Stieberger and Alexander Niemeyer for their assistance
during that time.
Z. L. was supported by the Alexander von Humboldt Fellowship.

%Figure 1:

%Figure 2:

%Figure 3:

%Figure 4:


\newpage
\begin{thebibliography}{99}
\newcommand{\np}{\mbox{\em {Nucl. Phys.} {\bf B }}}
\newcommand{\pl}{\mbox{\em {Phys. Lett.} {\bf }}}
\newcommand{\prl}{\mbox{\em {Phys. Rev. Lett. }}}
\newcommand{\cmp}{\mbox{\em {Comm. Math. Phys. }}}
\newcommand{\prd}{\mbox{\em {Phys. Rev.} {\bf D }}}
\newcommand{\pslondon}{\mbox{\em {Proc. R. Soc. London} {\bf A }}}
\bibitem{guth} A. H. Guth \prd {\bf 23} (1981) 347
\bibitem{rey} I. Yi, E. T. Vishniac and S. Mineshige
\prd {\bf 43 } (1991) 362, \\
I. Yi and E. T. Vishniac, \prd {\bf 47 } (1993) 5295, \\
S.--J. Rey, \np {\bf 284} (1987) 706
\bibitem{salopek2} D. S. Salopek and J. R. Bond, \prd {\bf 43} (1991) 1005
\bibitem{yi} I. Yi and E. T. Vishniac, \prd {\bf 47 } (1993) 5280
\bibitem{freese} K. Freese, J. A. Frieman and A. V. Olinto, \prl { \bf 65}
(1990) 3233
\bibitem{adams} F. C. Adams, J. R. Bond, K. Freese, J. A. Frieman and
A. V. Olinto, \prd {\bf 47} (1993) 426
\bibitem{risken} H. Risken, {\em The Fokker--Planck Equation,
Second Edition} (Springer--Verlag, Berlin, 1989)
\bibitem{recipes} W. H. Press, B. P. Flannery, S. A. Teukolsky and
W. T. Vetterling, {\em Numerical Recipes} (Cambridge University Press,
New York, 1987)
\bibitem{linde} A. Linde, {\em Particle Physics and Inflationary
Cosmology} (Harwood Academic Publishers, 1990)
\end{thebibliography}
\end{document}